\documentclass[11pt,twoside]{article}
\usepackage{hepro5}
\usepackage{graphicx}

\usepackage[T1]{fontenc} 

\usepackage{latexsym}
\usepackage{verbatim}

\begin{document}

\vskip 1.0cm
\markboth{L.~Foschini et al.}{Multiwavelength survey of RLNLS1s}
\pagestyle{myheadings}

\vspace*{0.5cm}
\title{Multiwavelength survey of a sample of flat-spectrum radio-loud narrow-line Seyfert 1 galaxies}

\author{L. Foschini$^1$,
M.~Berton$^2$,
A.~Caccianiga$^1$,
S.~Ciroi$^2$,
V.~Cracco$^2$,
B.~M. Peterson$^3$,
E.~Angelakis$^4$,
V.~Braito$^1$,
L.~Fuhrmann$^4$,
L.~Gallo$^5$,
D.~Grupe$^{6}$,
E.~J\"arvel\"a$^{7,8}$,
S.~Kaufmann$^{9,10}$,
S.~Komossa$^4$,
Y.~Y.~Kovalev$^{11,4}$,
A.~L\"ahteenm\"aki$^{7,8}$,
M.~M.~Lisakov$^{11}$,
M.~L. Lister$^{12}$,
S.~Mathur$^{3}$,
J.~L.~Richards$^{12}$,
P.~Romano$^{13}$,
A.~Sievers$^{14}$,
G.~Tagliaferri$^1$,
J.~Tammi$^7$,
O.~Tibolla$^{10,15}$,
M.~Tornikoski$^7$,
S.~Vercellone$^{13}$,
G.~La~Mura$^2$,
L.~Maraschi$^1$,
P.~Rafanelli$^2$}

\affil{$^1$INAF Osservatorio Astronomico di Brera, Merate (LC), Italy;\\
$^2$Dipartimento Fisica \& Astronomia Universit\`a di Padova, Padova, Italy;\\
$^3$Dept Astronomy \& Center for Cosmology \& AstroParticle Physics, The Ohio State University, Columbus, OH, USA;\\
$^4$Max-Planck-Institut f\"ur Radioastronomie, Bonn, Germany;\\
$^5$Dept Astronomy \& Physics, Saint Mary's University, Halifax, Canada;\\
$^6$Space Science Center, Morehead State University, Morehead, KY, USA;\\
$^7$Aalto University Mets\"ahovi Radio Observatory, Kylm\"al\"a, Finland;\\
$^8$Aalto University Dept Radio Science \& Engineering, Aalto, Finland;\\
$^{9}$Landessternwarte, Universit\"at Heidelberg, Heidelberg, Germany;\\
$^{10}$MCTP-UNACH, Tuxtla Guti\'errez, Chiapas, M\'exico.\\
$^{11}$Astro Space Center of the Lebedev Physical Institute, Moscow, Russia;\\
$^{12}$Dept Physics \& Astronomy, Purdue Univ., West Lafayette, IN, USA;\\
$^{13}$INAF Istituto di Astrofisica Spaziale e Fisica Cosmica, Palermo, Italy;\\
$^{14}$Institut de Radio Astronomie Millim\'etrique, Granada, Spain;\\
$^{15}$ITPA, Universit\"at W\"urzburg, W\"urzburg, Germany.}

\begin{abstract}
We report on a multiwavelength survey of a sample of 42 flat-spectrum radio-loud narrow-line Seyfert 1 galaxies (RLNLS1s). This is the largest known sample of this type of active galactic nucleus (AGN) to date. We found that 17\% of sources were detected at high-energy gamma rays ($E>100$~MeV), and 90\% at X-rays ($0.3-10$~keV). The masses of the central black holes are in the range $\sim 10^{6-8}M_{\odot}$, smaller than the values of blazars. The disk luminosities are about $1-49$\% of the Eddington value, with one outlier at 0.3\%, comparable with the luminosities observed in flat-spectrum radio quasars (FSRQs). The jet powers are $\sim 10^{42-46}$~erg~s$^{-1}$, comparable with BL Lac Objects, yet relatively smaller than FSRQs. However, once renormalized by the mass of the central black hole, the jet powers of RLNLS1s, BL Lacs, and FSRQs are consistent each other, indicating the scalability of the jets. We found episodes of extreme variability at high energies on time scales of hours. In some cases, dramatic spectral and flux changes are interpreted as the interplay between the relativistic jet and the accretion disk. We conclude that, despite the distinct observational properties, the central engines of RLNLS1s are similar to those of blazars. 
\end{abstract}

\section{The Survey}
As soon as the {\em Fermi} Large Area Telescope (LAT, Atwood et al. 2009) revealed high-energy $\gamma$-ray emission from a handful of radio-loud narrow-line Seyfert 1 galaxies (RLNLS1s, Abdo et al. 2009a,b,c, Foschini et al. 2010), we searched for more information about this type of AGN. Unlike their radio-quiet siblings, RLNLS1s were poorly observed and studied, probably because they are faint at all wavelengths. However, some pioneering surveys were carried out by Zhou \& Wang (2002), Komossa et al. (2006), Whalen et al. (2006), and Yuan et al. (2008), the latter comprising 23 RLNLS1s with radio loudness at 1.4~GHz greater than 100. After the {\em Fermi} discovery of high-energy $\gamma$~rays, Foschini (2011) performed a study on 76 NLS1s (46 radio loud, 30 radio quiet) based on archival data, where there was a clear lack of information in some frequency bands. In particular, only 60\% of sources were detected in the {\em ROSAT} All Sky Survey (X-rays). 

To optimize our efforts, we focussed on a subsample of RLNLS1s, mainly characterized by the flat or inverted radio spectral index ($\alpha_{\rm r} < 0.5$, $S_{\nu}\propto \nu^{-\alpha_{\rm r}}$), although we have taken into account also those sources with a single-frequency radio detection\footnote{We calculated the radio spectral index between $1.4$ and $5$~GHz when possible. When the latter measurement was missing, we used any other frequency available. Half of the sources (21/42) had only one detection at $1.4$~GHz. After the publication of our work, Richards \& Lister (2015) observed at 9~GHz one of these sources (J$0953+2836$) and calculated a spectral index of $\alpha_{1.4-9\,\rm{GHz}}=0.5$, which is still matching our requirements.}. In addition to what was available in public archives, we obtained new observations with the X-ray satellites {\em Swift} (25 sources) and {\em XMM-Newton} (3 sources). We studied also optical spectra from the Sloan Digital Sky Survey (SDSS), complemented with new observations (2 sources) from the Asiago Astrophysical Observatory (Italy). Here we present an extended summary of the paper by Foschini et al. (2015). In addition, extensive multifrequency radio observations with high sampling rate of four $\gamma$-detected RLNLS1s are reported by Angelakis et al. (2015), while L\"ahteenm\"aki et al. (in preparation) is performing a high-frequency ($22-37$~GHz) radio survey. Infrared properties ({\em WISE}) have been analysed in detail by Caccianiga et al. (2015). The parent population, which comprises the sources with the jet viewed at large angles, was studied by Berton et al. (2015a).

\begin{figure}
\begin{center}
\hspace{0.25cm}
\includegraphics[angle=270,scale=0.42]{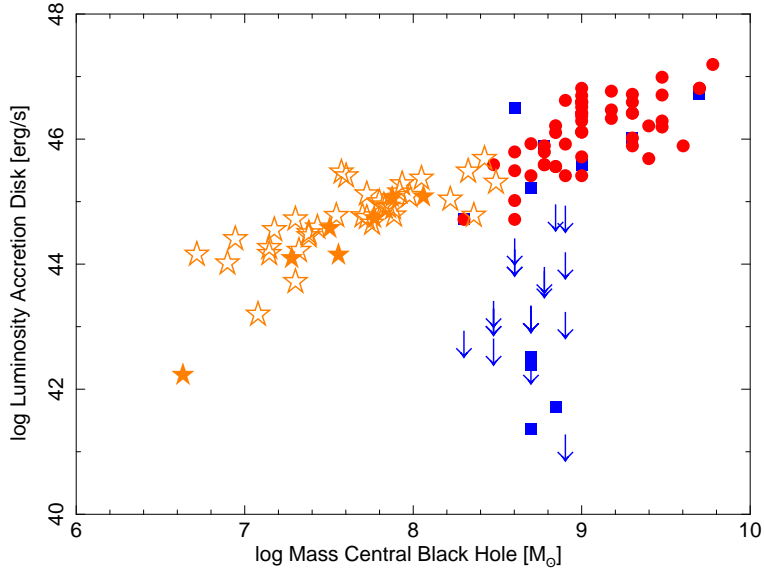}

\caption{The parameter space of the luminosity of the accretion disk and the mass of the central black hole for AGN with powerful relativistic jets. FSRQs (red filled circles), BL Lacs (blue filled squares, blue arrows for upper limits), RLNLS1s (orange stars; $\gamma$-ray detections are indicated with filled orange stars).}
\label{massdisk}
\end{center}
\end{figure}

The distributions of masses of the central black holes and the luminosities of the accretion disks are shown in Fig.~\ref{massdisk}, together with a sample of blazars (FSRQs and BL Lac Objects) for comparison. It is evident that RLNLS1s cover a previously unexplored part of that parameter space. This is crucial in the unification of powerful relativistic jets from stellar-mass black holes and neutron stars to AGN, as shown by Foschini (2014).

We found some interesting observational characteristics of RLNLS1s. About 17\% of the sources (7/42) were detected at MeV--GeV energies\footnote{After the publication of our work, D'Ammando et al. (2015) reported about the $\gamma$-ray detection of one source of our sample, J$1644+2619$, thus increasing the detection rate to 19\% (8/42). Three new $\gamma$-ray detections of RLNLS1s -- not included in the present sample -- were reported by Komossa et al. (2015), Liao et al. (2015), Yao et al. (2015).}, with an average spectral index $\alpha_{\gamma}\sim 1.6$, similar to FSRQs ($\sim 1.4$). We detected about 90\% of the sources (38/42) at X-rays (keV energies), but 3/4 of the undetected objects were never observed at X-rays and we found an upper limit in the RASS only. The X-ray spectral index of RLNLS1s is similar to broad-line Seyfert 1s ($\alpha_{\rm X,RLNLS1}\sim 1.0$, $\alpha_{\rm X,BLS1}\sim 1.1$) and in the middle between FSRQs and BL Lac Objects ($\alpha_{\rm X,FSRQ}\sim 0.6$, $\alpha_{\rm X,BL\,Lac}\sim 1.3$). The spectral energy distributions (SEDs) of RLNLS1s showed that the X-ray emission could be due to inverse-Compton emission either from the relativistic jet (non thermal) or the corona of the accretion disk (thermal). This could explain why the average spectral index is softer than that of FSRQs, where the X-ray emission is dominated by the jet only. Infrared colours are typical of synchrotron emission, although there is a significant overlap with the starburst region in the {\em WISE} color-color diagram (see also Caccianiga et al. 2015).

Violent intraday variability -- even on hour timescales -- has been observed in many sources in the sample at all wavelengths. In some cases, where a dense  multiwavelength coverage was available, we observed strong spectral changes in the SED that could be explained by the jet-disk interplay. Radio measurements indicate changes only in the Very Large Baseline Interferometer (VLBI) core, and in a few cases, when compared to $\gamma$-ray lightcurves, suggested a link between the rotation of the electric vector position angle (EVPA) and the emission at high energy. 

The calculated jet power is in the range $10^{42.6-45.6}$~erg~s$^{-1}$, generally lower than blazars, yet consistent if normalised by the mass of the central black hole. We conclude that RLNLS1s are similar to other AGN with a powerful relativistic jets. Specifically, they seem to be the low-power tail of the FSRQ distribution. The small number of sources known today is likely due to either their low observed luminosities or an on-off behaviour of the jet. Future observatories with better sensitivities (e.g. SKA, Berton et al. 2015b), should reveal more sources of this type. 

\acknowledgments {\em Swift} observations were partially supported by the contract ASI-INAF I/004/11/0. The Mets\"ahovi team acknowledges the support from the Academy of Finland (numbers 212656, 210338, 121148, and others). YYK and MML are partly supported by the Russian Foundation for Basic Research (project 13-02-12103). Y.Y.K.\ is also supported by the Dynasty Foundation. BMP is supported by the NSF through grant AST-1008882. This research has made use of data from the MOJAVE database that is maintained by the MOJAVE team (Lister et al. 2009, 2013) and supported under NASA {\em Fermi} grant NNX12AO87G. JLR acknowledges support from NASA through {\em Fermi} Guest Investigator grant NNX13AO79G.

\end{document}